\documentclass[conference]{IEEEtran}
\IEEEoverridecommandlockouts
\usepackage{cite}
\usepackage{amsmath,amssymb,amsfonts}
 \usepackage{verbatim}
\usepackage{graphicx}
 \usepackage{url} 
\usepackage{mdframed}
\usepackage{multirow}
\usepackage{booktabs}
\usepackage[table,xcdraw]{xcolor}
\usepackage{textcomp}
\usepackage{xcolor}
\usepackage{algpseudocode}
\usepackage{algorithm}
\usepackage[table]{xcolor} % for \rowcolor
\usepackage{multirow}      % for \multirow
\usepackage{booktabs}      % for \toprule, \midrule, \bottomrule
\usepackage[utf8]{inputenc}
\usepackage{tcolorbox}
\tcbuselibrary{skins}

\def\BibTeX{{\rm B\kern-.05em{\sc i\kern-.025em b}\kern-.08em
    T\kern-.1667em\lower.7ex\hbox{E}\kern-.125emX}}
\begin{document}

\title{Toward Green Code: Prompting Small Language Models for Energy-Efficient Code Generation}
\author{
\IEEEauthorblockN{Humza Ashraf\IEEEauthorrefmark{1}, Syed Muhammad Danish\IEEEauthorrefmark{1}, Shadikur Rahman\IEEEauthorrefmark{1}, Zeeshan Sattar\IEEEauthorrefmark{2}} 
\IEEEauthorblockA{\IEEEauthorrefmark{1}Algoma University, Brampton, Canada\\
\IEEEauthorrefmark{2}Ericsson Inc., Ottawa, Canada\\
Emails: \{hashraf, syed.danish, shadikur.rahman\}@algomau.ca, zeeshan.sattar@ericsson.com}
}

\maketitle

\begin{abstract}

%Large Language Models (LLMs) are increasingly used for code generation and completion tasks, helping developers save time and increase productivity. Despite their strong ability to generate functionally correct code, commercially available LLMs, such as ChatGPT, require a large amount of computing power, leading to increased energy consumption and carbon emissions. This has raised concerns about their environmental impact. 

There is a growing concern about the environmental impact of large language models (LLMs) in software development, particularly due to their high energy use and carbon footprint. Small Language Models (SLMs) offer a more sustainable alternative, requiring fewer computational resources while remaining effective for fundamental programming tasks. In this study, we investigate whether prompt engineering can improve the energy efficiency of SLMs in code generation. We evaluate four open-source SLMs, StableCode-Instruct-3B, Qwen2.5-Coder-3B-Instruct, CodeLlama-7B-Instruct, and Phi-3-Mini-4K-Instruct, across 150 Python problems from LeetCode, evenly distributed into easy, medium, and hard categories. Each model is tested under four prompting strategies: role prompting, zero-shot, few-shot, and chain-of-thought (CoT). For every generated solution, we measure runtime, memory usage, and energy consumption, comparing the results with a human-written baseline. Our findings show that CoT prompting provides consistent energy savings for Qwen2.5-Coder and StableCode-3B, while CodeLlama-7B and Phi-3-Mini-4K fail to outperform the baseline under any prompting strategy. These results highlight that the benefits of prompting are model-dependent and that carefully designed prompts can guide SLMs toward greener software development.
\end{abstract}


\section{Introduction}

Large Language Models (LLMs) \cite{rahman2025refactorcoderqa} have achieved remarkable success in code generation \cite{herrington2003code}. In both research and industry, several advanced models have been developed, including GitHub Copilot, CodeLlama, and ChatGPT. The accuracy of the code produced by these models is now comparable to that of human developers. However, despite these advances, the environmental impact of LLMs remains substantial, with training alone resulting in significant CO${2}$ emissions and high water consumption \cite{ashraf2025energy}. For instance, training LLaMA 3.1 (8B parameters) produced approximately 420 tCO${2}$e, equivalent to 83 years of electricity usage by a single U.S. household \cite{morrison2503holistically}. As demand for AI services increases, the electricity consumption of data centers that run these models also rises. By 2030, data center power demand is projected to grow by 160\%, with AI expected to make up nearly 19\% of this demand by 2028 \cite{de2025ai}. This growth not only raises operational costs but also creates sustainability challenges, particularly when energy comes from non-renewable sources. Consequently, LLM-based applications are facing critical questions regarding their long-term sustainability.

Given these concerns, Small Language Models (SLMs) offer a more sustainable path forward. With fewer parameters and simpler architectures, SLMs require less computational power, memory, and energy for both training and inference. They are well-suited for simpler tasks such as solving fundamental coding problems, where efficiency and speed are prioritized \cite{chen2024role}. Moreover, their lightweight design makes them highly suitable for deployment in edge environments, where resources are limited and low-latency processing is critical \cite{li2025collaborative}. While SLMs are capable of addressing moderately complex programming challenges, they often struggle with long-term contexts or deeply structured code. The difference highlights potential pathways for exploring SLMs that are both efficient and sustainable while maintaining performance. If adopted more widely, SLMs could help reduce the overall energy demand of AI applications and support greener software development.

Prompt engineering \cite{gao2023prompt} has recently emerged as a powerful technique for improving the performance of language models without increasing their size or computational cost. By carefully designing input prompts, users can guide models to produce more accurate, efficient, and context-aware outputs. This approach is particularly relevant to code generation, where small changes in prompts can significantly influence the correctness, readability, and energy efficiency of the generated code. Previous studies \cite{peng2024large, rubei2025prompt, tuttle2024can, cappendijk2025exploration, wang2024advanced, niu2024evaluating, waghjale2024ecco, podder2025empirical, jonnala2025measuring, hou2025comparing} have applied prompt engineering to promote sustainable code generation {citations}. However, most of this work focuses on LLMs and overlooks SLMs, where efficiency gains could be even more valuable. This raises an important research question (RQ): \textit{Can prompt engineering help SLMs generate energy-efficient code for sustainable software development?}
\begin{figure*}[t!]
    \centering
    \includegraphics[width=\linewidth]{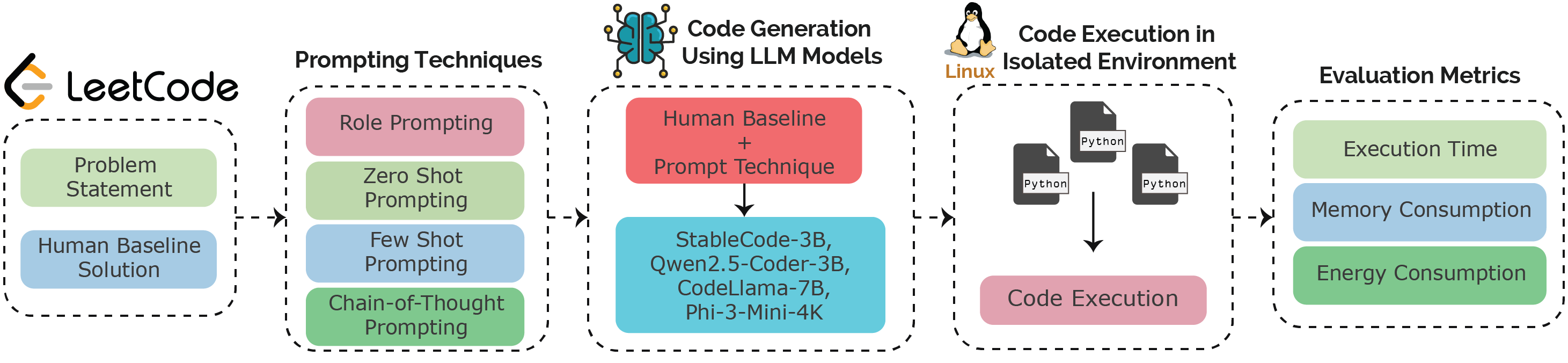}
    \caption{Overall Methodology}
    \label{fig:model}
\end{figure*}

In this work, we address this question by applying different prompt engineering techniques to SLMs. We hypothesize that carefully designed sustainability-oriented prompts can guide SLMs to generate more efficient code. The objective of this study is to identify prompt engineering strategies that improve the sustainability of code generation without increasing model size or computational demands. Importantly, we do not measure the energy consumed by running the language models themselves. Instead, our analysis focuses on the energy efficiency of the code produced by SLMs under various prompt engineering approaches. 

To answer the \textit{RQ} presented above, this work makes the following key contributions:

\begin{itemize}
    \item We evaluate four SLMs, StableCode-Instruct-3B, Qwen2.5-Coder-3B-Instruct, CodeLlama-7B-Instruct, and Phi-3-Mini-4K-Instruct, using four prompt engineering strategies: role prompting, zero-shot, few-shot, and chain-of-thought (CoT). Unlike prior studies, our focus is on assessing whether sustainability-oriented prompts can guide SLMs to generate green code.
    \item Our experiments apply four different prompting strategies to 150 Python problems from LeetCode, covering equal numbers of easy, medium, and hard tasks. In an isolated Linux environment, we evaluate the performance and efficiency of SLMs by measuring runtime, memory usage, and energy consumption for each generated solution, and then compare these results against a human-written code baseline.
\end{itemize}
%We conduct experiments by prompting LLMs to generate solutions for programming problems and then measure various statistics, including energy consumption, during the execution of the generated code in an isolated Linux environment. In this study, code generation is evaluated across three levels of difficulty using the Python programming language. In total, 150 coding problems, 50 of each difficulty level (easy, medium, and hard), are taken from LeetCode \cite{leetcode}, a platform for assessing the programming capabilities of LLMs. Our comparison includes three small open-source models, StableCode-3B, StarCoderBase-3B, and Qwen2.5-Coder-3B-Instruct, and two large commercial models, GPT-4.0 and DeepSeek-Reasoner, accessed via paid APIs. The primary goal is to compare run-time performance, energy consumption, and memory consumption of the generated code. 

Results show that SLMs such as Qwen2.5-Coder and StableCode-3B achieve the best energy efficiency, with CoT prompting consistently outperforming the baseline across various difficulty levels. In contrast, models like CodeLlama-7B and Phi-3-Mini-4K did not perform well, as none of their prompting strategies were able to reduce energy consumption below the baseline.

%The rest of the paper is organized as follows. Section \ref{sec:rl} presents the related work. Section \ref{sec:meth} presents the overall methodology. Sec. \ref{sec:result} discusses the analysis and results. Section \ref{sec:ttv} presents the limitations, while Section \ref{sec:con} concludes the paper.

\section{Related Work}
\label{sec:rl}
Peng et al. \cite{peng2024large} proposed an automated LLM-assisted tool for software refactoring that improves energy efficiency while preserving program semantics. Their approach uses prompt engineering and feedback loops with energy profiling, enabling the LLM to optimize code for both speed and lower power consumption. Rubei et al. \cite{rubei2025prompt} studied how prompt engineering can reduce the carbon footprint of LLM-based code generation. Using the Llama 3 model on the CodeXGLUE benchmark, they evaluated both energy consumption and accuracy. Tuttle et al. \cite{tuttle2024can} examined the role of prompt engineering in energy-efficient code generation through a three-step prompting methodology. Their results show that while prompt strategies affect runtime and energy use, models remain inconsistent and often perform worse than human-written solutions. Cappendijk et al. \cite{cappendijk2025exploration} explored zero-shot and few-shot prompting for energy-efficient code generation. Results indicate that certain combinations of prompts, models, and problems reduce energy usage, but no strategy is consistently effective. Wang et al. \cite{wang2024advanced} evaluated prompt engineering in advanced LLMs by comparing the non-reasoning GPT-4o with the reasoning model o1 on software engineering tasks such as code generation, translation, and summarization.

Niu et al. \cite{niu2024evaluating} analyzed the efficiency of LLM-generated code, moving beyond correctness. They tested different prompts and found that simple prompts improve efficiency for basic problems, while chain-of-thought prompting benefits more complex ones. Waghjale et al. \cite{waghjale2024ecco} studied code generation approaches such as instruction prompting and few-shot learning to improve program efficiency while preserving correctness. Their findings show no single model consistently leads in efficiency, highlighting trade-offs between accuracy and energy/runtime performance. Podder et al. \cite{podder2025empirical} investigated whether prompt engineering can make AI coding assistants like GitHub Copilot produce greener code. They developed a prompt template and optimization strategy, applying it to Java programs and an enterprise application. Their method reduced energy use and carbon emissions by about 36\%, showing that prompt engineering can support sustainable software development.

Jonnala et al. \cite{jonnala2025measuring} evaluated Python code generated by GPT-4o-Mini, GPT-3.5-Turbo, and GPT-4-Turbo, measuring execution time and memory use while ensuring correctness. Using chain-of-thought prompting, they found efficiency gains for GPT-4o-Mini and GPT-3.5-Turbo but not for GPT-4-Turbo. Finally, Hou et al. \cite{hou2025comparing} compared seven LLMs for code generation across prompt strategies, programming languages, and task difficulties. They identified GPT-4 as the strongest overall, though its performance varied significantly with different prompting strategies.

\subsection{Novelty}
While previous works \cite{peng2024large, rubei2025prompt, tuttle2024can, cappendijk2025exploration, wang2024advanced, niu2024evaluating, waghjale2024ecco, podder2025empirical, jonnala2025measuring, hou2025comparing} have primarily examined prompt engineering and the energy impact of large LLMs, little attention has been given to SLMs. In this study, we investigate whether prompt engineering can enhance the energy efficiency of SLMs in code generation. We consider different prompt engineering techniques, including zero-shot, few-shot, and chain-of-thought prompting, to assess their effectiveness across problems of varying algorithmic complexity. This study will help researchers and developers by filling an important gap in understanding the role of SLMs in energy-efficient code generation. It identifies which prompt engineering strategies work best for specific models and problem complexities, offering clear guidance on prompt selection.

\section{Methodology}
\label{sec:meth}
We evaluate and compare the energy consumption and performance of code generated by different SLMs using sustainability-focused prompts, with human-written solutions serving as the efficiency baseline. The overall methodology of the proposed work is illustrated in Fig. \ref{fig:model}. Following the formulation proposed by Basili et al. \cite{caldiera1994goal}, our high-level goal can be summarized in the following primary research question:

\begin{mdframed}
\textit{Can prompt engineering help SLMs generate energy-efficient code for sustainable software development?}
\end{mdframed}

We hypothesize that carefully designed sustainability-oriented prompts can guide SLMs to produce more efficient code. The aim of this study is to compare the outputs of different SLMs against each other and against baseline human-written solutions.

\subsection{Selection of Dataset and Baseline}
In this work, we focus on Python because it is widely used in both education and industry. For our experiments, we selected coding problems from LeetCode, which organizes challenges into three levels of difficulty: \textit{easy}, \textit{medium}, and \textit{hard}. We randomly chose 150 problems in total, with 50 from each category, ensuring that all had community-verified solutions. As a baseline, we used human-written solutions for programming problems available on LeetCode. Prior research has widely adopted LeetCode solutions \cite{niu2024evaluating, solovyeva2025ai} as benchmarks because the platform provides a broad range of problems and ranks solutions through community voting. This makes it a reliable source for identifying high-quality code written by experienced developers. For each of the 150 problems in our study, we selected one Python solution with the highest number of community up-votes, specifically those recognized for clarity and optimized time and space complexity. These solutions served both as baseline references for comparison and as prompts to evaluate whether SLMs could generate more efficient and sustainable code.

\subsection{Selection of LLMs}
In this study, we used four instruction-tuned SLMs: StableCode-3B \cite{stablecode3b}, Qwen2.5-Coder-3B \cite{qwen25coder3b}, CodeLlama-7B \cite{codellama7b}, and Phi-3-Mini-4K \cite{phi3mini4k}, designed for efficient reasoning and instruction following. We selected \textit{instruction-tuned} models for this study because they are specifically optimized to follow natural language instructions, making them well-suited for prompt engineering experiments. These models are trained on datasets that align code generation with user-provided instructions, ensuring that variations in prompt design have a direct and measurable effect on the generated output. This characteristic makes them particularly appropriate for evaluating the impact of sustainability-focused and natural language prompts on code generation. To reduce space, we omit the term \textit{Instruct} from the names of the models throughout the paper.

\subsection{Prompting Strategies}
We used four prompting techniques in this study: role prompting, zero-shot, few-shot, and CoT. For zero-shot and CoT prompting, we adopted the prompt templates introduced by \cite{niu2024evaluating, shypula2023learning}. In each case, a human-written solution from LeetCode was provided to the SLM along with the corresponding prompt, instructing the model to optimize the code.

\subsubsection{Role Prompting} In role prompting \cite{wang2024role}, the model is assigned a specific role to guide its reasoning and output style. We instructed the SLMs to act as a senior software engineer with 10 years of experience in designing and optimizing software systems. The models were then asked to carefully review the provided human-written LeetCode solution and rewrite it in a way that improves time and space efficiency while maintaining correctness. The exact prompt is given below.

\begin{tcolorbox}[
    colback=gray!5,          % background color
    colframe=black!70,       % border color
    title=Role Prompting,          % box title
    fonttitle=\bfseries,     % bold title
    boxrule=0.6pt,           % border thickness
    arc=3mm,                 % <-- round edges
    enhanced,                % better rendering
    sharp corners=downhill,  % smoother bottom edges
    width=\linewidth      % adjust width
]
\textbf{User:} \{Human Baseline Solution\} \\
You are a senior software engineer with 10 years of experience in designing and optimizing software systems. Carefully review the given code and rewrite it to be as energy-efficient and optimized as possible. Focus on improving time and space complexity while ensuring correctness. Provide only the optimized code in your output. \\[6pt]
\textbf{SLM:} \{Output Code\}
\end{tcolorbox}

\subsubsection{Zero-Shot Prompting}In zero-shot prompting, the model receives the task without any additional examples. Each SLM was provided with a LeetCode human-written solution along with a direct instruction to optimize the code for efficiency. No further context or demonstrations were given. The exact prompt is given below.

\begin{tcolorbox}[
    colback=gray!5,          % background color
    colframe=black!70,       % border color
    title=Zero-Shot Prompting,          % box title
    fonttitle=\bfseries,     % bold title
    boxrule=0.6pt,           % border thickness
    arc=3mm,                 % <-- round edges
    enhanced,                % better rendering
    sharp corners=downhill,  % smoother bottom edges
    width=\linewidth      % adjust width
]
\textbf{User:} \{Human Baseline Solution\} \\
Please optimize the following code to make it as \textit{energy-efficient} as possible. Focus on reducing both time and space complexity while preserving correctness. Provide only the optimized code as output. \\[6pt]
\textbf{SLM:} \{Output Code\}
\end{tcolorbox}

\subsubsection{Few-Shot Prompting}
Few-shot prompting provides the model with one or more worked examples before the actual task. For this setup, we constructed prompts containing a small set of code optimization examples, each showing an unoptimized piece of code followed by its optimized version. After these examples, the target LeetCode solution was provided to the model, with the instruction to optimize it in a similar manner. The exact prompt is given below.

\begin{tcolorbox}[
    colback=gray!5,          % background color
    colframe=black!70,       % border color
    title=Few-Shot Prompting,          % box title
    fonttitle=\bfseries,     % bold title
    boxrule=0.6pt,           % border thickness
    arc=3mm,                 % <-- round edges
    enhanced,                % better rendering
    sharp corners=downhill,  % smoother bottom edges
    width=\linewidth      % adjust width
]
\textbf{User:} Below are examples of unoptimized and optimized code. Learn the pattern of optimization and then optimize the provided solution to make it more \textit{energy-efficient}, with reduced time and space complexity. Provide only the optimized code as output. \\

\textit{Example 1 (Unoptimized):}
{Unoptimized code snippet}

\textit{Example 1 (Optimized):}
{Optimized code snippet}

\textit{Example 2 (Unoptimized):}
{Unoptimized code snippet}

\textit{Example 2 (Optimized):}
{Optimized code snippet}

\textit{Target Code to Optimize:}
{Human Baseline solution} \\

\textbf{SLM:} \{Output Code\}
\end{tcolorbox}

\subsubsection{CoT Prompting}
In CoT prompting, the model is encouraged to reason through the task step by step before producing an optimized solution. In our setup, we first provided the human-written LeetCode solution to the GPT-5 API and asked it to outline a strategy for optimizing the code. We used GPT-5 because of its strong reasoning capabilities, which allowed it to identify performance bottlenecks and suggest concrete improvements in terms of time and space complexity. We then provided both the strategy and the original human-written solution to the SLMs, instructing them to optimize the code using the given reasoning steps. The exact prompt is given below.

\begin{tcolorbox}[
    colback=gray!5,          % background color
    colframe=black!70,       % border color
    title=CoT Prompting,          % box title
    fonttitle=\bfseries,     % bold title
    boxrule=0.6pt,           % border thickness
    arc=3mm,                 % <-- round edges
    enhanced,                % better rendering
    sharp corners=downhill,  % smoother bottom edges
    width=\linewidth      % adjust width
]

\textbf{User:} \{Human Baseline solution\} \\
Analyze the above code and provide a potential strategy to improve its \textit{energy efficiency}, focusing on reducing time and space complexity while maintaining correctness. \\
\textbf{GPT-5 API:} Strategy \\

\textbf{User:} Now apply the strategy mentioned above and provide the optimized version of the same code. Return only the optimized code as output. \\
\textbf{SLM:} \{Output Code\} 

\end{tcolorbox}
\begin{table*}[ht!]
\centering
\caption{Performance comparison of LLMs under different prompting strategies, averaged across easy, medium, and hard problems (50 each, including the baseline)}
\renewcommand{\arraystretch}{1.15}
\scriptsize

\begin{tabular}{l c ccc ccc ccc ccc}
\toprule
\textbf{Model} & \textbf{Category} &
\multicolumn{3}{c}{\textbf{CoT}} &
\multicolumn{3}{c}{\textbf{Few Shot}} &
\multicolumn{3}{c}{\textbf{Zero Shot}} &
\multicolumn{3}{c}{\textbf{Role}} \\
\cmidrule(lr){3-5} \cmidrule(lr){6-8} \cmidrule(lr){9-11} \cmidrule(lr){12-14}
& & Runtime & Memory & Energy & Runtime & Memory & Energy & Runtime & Memory & Energy & Runtime & Memory & Energy \\
\midrule
\multirow{3}{*}{Qwen2.5-Coder} 
 & Easy   & 0.00557 & 614.35 & 1.7103 & 0.00608 & 643.06 & 1.7120 & 0.00600 & 640.79 & 1.7112 & 0.00599 & 640.79 & 1.7112 \\
 & Medium & 0.00584 & 637.98 & 1.7111 & 0.00634 & 665.22 & 1.7131 & 0.00614 & 654.54 & 1.7129 & 0.00612 & 654.54 & 1.7114 \\
 & Hard   & 0.00610 & 664.48 & 1.7125 & 0.00609 & 647.63 & 1.7114 & 0.00595 & 640.60 & 1.7107 & 0.00597 & 640.60 & 1.7119 \\
\midrule
\multirow{3}{*}{CodeLlama-7B}
 & Easy   & 0.00580 & 626.30 & 1.7149 & 0.00582 & 622.25 & 1.7185 & 0.00596 & 632.38 & 1.7117 & 0.00593 & 632.39 & 1.7120 \\
 & Medium & 0.00618 & 654.93 & 1.7128 & 0.00630 & 666.75 & 1.7144 & 0.00623 & 660.43 & 1.7127 & 0.00624 & 660.26 & 1.7131 \\
 & Hard   & 0.00587 & 635.99 & 1.7118 & 0.00563 & 606.57 & 1.7103 & 0.00605 & 646.26 & 1.7127 & 0.00606 & 646.28 & 1.7119 \\
\midrule
\multirow{3}{*}{StableCode-3B}
 & Easy   & 0.00588 & 653.74 & 1.7124 & 0.00601 & 637.21 & 1.7126 & 0.00589 & 626.20 & 1.7126 & 0.00589 & 626.19 & 1.7125 \\
 & Medium & 0.00574 & 633.23 & 1.7117 & 0.00626 & 659.34 & 1.7136 & 0.00610 & 648.92 & 1.7130 & 0.00611 & 648.91 & 1.7143 \\
 & Hard   & 0.00539 & 611.99 & 1.7099 & 0.00602 & 642.66 & 1.7140 & 0.00650 & 662.79 & 1.7156 & 0.00649 & 662.62 & 1.7144 \\
\midrule
\multirow{3}{*}{Phi-3-Mini-4K}
 & Easy   & 0.00608 & 641.28 & 1.7132 & 0.00600 & 633.75 & 1.7131 & 0.00508 & 571.53 & 1.7100 & 0.00567 & 629.45 & 1.7112 \\
 & Medium & 0.00757 & 666.12 & 1.7175 & 0.00625 & 651.86 & 1.7144 & 0.00576 & 615.20 & 1.7168 & 0.00610 & 641.58 & 1.7150 \\
 & Hard   & 0.00607 & 648.63 & 1.7130 & 0.00612 & 644.76 & 1.7143 & 0.00576 & 637.15 & 1.7114 & 0.00604 & 650.78 & 1.7133 \\
\midrule
\multirow{3}{*}{Baseline}
 & Easy   & 0.00598 & 641.11 & 1.7114 & 0.00598 & 641.11 & 1.7114 & 0.00598 & 641.11 & 1.7114 & 0.00598 & 641.11 & 1.7114 \\
 & Medium & 0.00620 & 659.81 & 1.7138 & 0.00620 & 659.81 & 1.7138 & 0.00620 & 659.81 & 1.7138 & 0.00620 & 659.81 & 1.7138 \\
 & Hard   & 0.00602 & 644.72 & 1.7115 & 0.00602 & 644.72 & 1.7115 & 0.00602 & 644.72 & 1.7115 & 0.00602 & 644.72 & 1.7115 \\
\bottomrule
\end{tabular}
\label{tab:1}
\end{table*}

\begin{table*}[ht!]
\centering

\caption{Performance comparison of LLMs under different prompting strategies, averaged across all 150 problems (bold indicates the best or lowest value per column, including the baseline)}
\renewcommand{\arraystretch}{1.15}
\scriptsize
\begin{tabular}{l ccc ccc ccc ccc}
\toprule
\textbf{LLM Model} &
\multicolumn{3}{c}{\textbf{CoT}} &
\multicolumn{3}{c}{\textbf{Few Shot}} &
\multicolumn{3}{c}{\textbf{Zero Shot}} &
\multicolumn{3}{c}{\textbf{Role}} \\
\cmidrule(lr){2-4} \cmidrule(lr){5-7} \cmidrule(lr){8-10} \cmidrule(lr){11-13}
& Runtime & Memory & Energy & Runtime & Memory & Energy & Runtime & Memory & Energy & Runtime & Memory & Energy \\
\midrule
Qwen2.5-Coder-3B         & 0.00584 & 638.76 & \textbf{1.7112} & 0.00617 & 651.99 & \textbf{1.7121} & 0.00603 & 645.33 & \textbf{1.7115} & 0.00603 & 645.33 & \textbf{1.7114} \\
CodeLlama-7B          & 0.00595 & 639.09 & 1.7131 & \textbf{0.00592} & \textbf{631.85} & 1.7144 & 0.00608 & 646.35 & 1.7124 & 0.00607 & 646.31 & 1.7123 \\
StableCode-3B         & \textbf{0.00567} & \textbf{633.13} & 1.7113 & 0.00610 & 646.40 & 1.7133 & 0.00616 & 645.85 & 1.7137 & 0.00616 & 645.79 & 1.7137 \\
Phi-3-Mini-4K         & 0.00656 & 651.91 & 1.7145 & 0.00613 & 643.45 & 1.7139 & \textbf{0.00553} & \textbf{607.76} & 1.7127 & \textbf{0.00594} & \textbf{640.60} & 1.7131 \\
\midrule
Baseline              & 0.00606 & 648.57 & 1.7122 & 0.00606 & 648.57 & 1.7122 & 0.00606 & 648.57 & 1.7122 & 0.00606 & 648.57 & 1.7122 \\
\bottomrule
\end{tabular}
\label{tab:2}
\end{table*}
\subsection{Code Generation} After defining the prompting techniques, the next step was to generate code with the SLMs. All four models received the same prompts for each prompting technique, and experiments were conducted independently for each case. The models used in this study are open-source and were downloaded locally through Hugging Face. We developed automated scripts that applied the specified prompt engineering technique to call each model and generate code, which was then saved as individual Python files for further processing. For each of the 150 problems, one model produced 150 Python solutions per prompting technique. Since four prompting techniques were applied, this resulted in 150 × 4 = 600 solutions per model. In total, 2400 Python codes were generated across all four SLMs and techniques. In some instances, SLMs produced multiple solutions or echoed the provided human-written baseline solution and problem statement. To address this, we performed a code-cleaning process using a combination of manual review and the ChatGPT API, retaining only the first valid solution. This ensured that all models were evaluated fairly, with exactly one solution per problem. All code generation and execution were performed using Python 3.

\subsection{Sustainability Metrics}
In this study, we evaluate the efficiency and environmental impact of the generated code using three key metrics: runtime, memory usage, and energy consumption.

\subsubsection{Run-time}
Runtime refers to the time taken by the code to execute and return a result, and in this study it is measured in milliseconds (ms). To capture runtime, we used Python’s built-in \texttt{time} module, which provides a simple and effective way to measure the execution duration of scripts or code blocks.
\subsubsection{Memory Consumption}
During execution, the code consumes memory, with the highest point recorded as its peak memory usage. In this study, memory usage is measured in kibibytes (KiB). To capture peak memory usage, we used Python’s built-in \texttt{tracemalloc} module. For each code sample, memory tracing was activated with \texttt{tracemalloc.start()}, and the peak value was retrieved after execution using \texttt{tracemalloc.get\_traced\_memory()}, which reports both current and peak memory usage.
\subsubsection{Energy Consumption}
Energy consumption refers to the total amount of energy used by the code during execution, primarily reflecting CPU usage. Lower energy consumption indicates greater efficiency and environmental sustainability. In this study, energy usage is measured in milliwatt-hours (mWh). To estimate energy consumption, we used the \texttt{CodeCarbon} Python library, which monitors CPU activity to track the energy footprint of Python code.

\section{Results}
\label{sec:result}
%In this section, we present and analyze the evaluation results to answer \textbf{RQ}.

\subsection{Measurement Environment and Experimental Setup} 
To generate the code, we used an NVIDIA A100 40GB GPU on Google Cloud, where the models were downloaded locally and loaded into memory. All experiments were conducted on Linux in an isolated environment, where Python code generated by the SLMs was executed and analyzed. For consistency and control, the experiments were carried out on a Google Cloud Compute Engine VM of type \texttt{c2-standard-8}, located in the \texttt{us-central1-c} region. The VM runs Ubuntu 24.04 LTS on a 100 GB SSD with an \texttt{x86\_64} architecture. A C2 instance was chosen because it provides dedicated CPU cores and stable performance, which is essential for reproducible measurements. All scripts were executed using Python 3.12.3.

Each code sample was executed ten times to account for variability and ensure reliable results. A five-second cooling interval was applied between runs to maintain stable conditions. To minimize external or nondeterministic influences, all executions were performed under identical virtualized settings.

\subsection{Performance Comparison of SLMs Under Different Prompting Strategies}

Table \ref{tab:1} compares the performance of four SLMs (Qwen2.5-Coder-3B, CodeLlama-7B, StableCode-3B, and Phi-3-Mini-4K) under four prompting strategies, across Easy, Medium, and Hard tasks. For each setting, the table reports runtime, memory usage, and energy consumption, with a baseline configuration included for reference. Overall, the results indicate that all models consume nearly the same amount of energy, typically between 1.71 and 1.715 mWh, while runtime and memory usage show more variation. Qwen2.5-Coder-3B and StableCode-3B generally achieve lower runtimes and memory consumption, suggesting higher efficiency in certain tasks. In contrast, CodeLlama-7B and Phi-3-Mini-4K exhibit greater fluctuations, particularly for Medium tasks where runtimes are higher and energy consumption slightly increases. 

Table \ref{tab:2} compares four LLMs under different prompting strategies and shows clear differences in runtime, memory, and energy efficiency. Runtime varies the most: StableCode-3B is generally the fastest with CoT prompting, Phi-3-Mini-4K runs quicker with Zero-Shot and Role prompting, and CodeLlama-7B performs best in Few-Shot tasks. This shows that runtime efficiency depends on both the model and the prompting style, and no single model is the fastest in every case. Memory usage also shows noticeable differences. Phi-3-Mini-4K uses the least memory in Zero-Shot prompting, which makes it suitable for situations where resources are limited, while StableCode-3B is efficient under CoT. In contrast, Qwen2.5-Coder-3B and CodeLlama-7B generally use more memory, showing that they are less affected by changes in prompting style. %Energy consumption is the most stable across models and prompting methods, with all results staying close to the baseline. Qwen2.5-Coder-3B is slightly more efficient in energy use, while the other models often match or exceed the baseline.

\begin{tcolorbox}[colback=gray!10, colframe=black, boxrule=0.3mm, arc=0mm, left=2mm, right=2mm, top=1mm, bottom=1mm]
\textbf{Summary:} The results reveal that no model excels in all scenarios, with some underperforming the baseline. Model performance is highly dependent on the prompting strategy, which can either enhance or reduce efficiency.
\end{tcolorbox}

\begin{figure*}[t!]
    \centering
    \includegraphics[width=\linewidth]{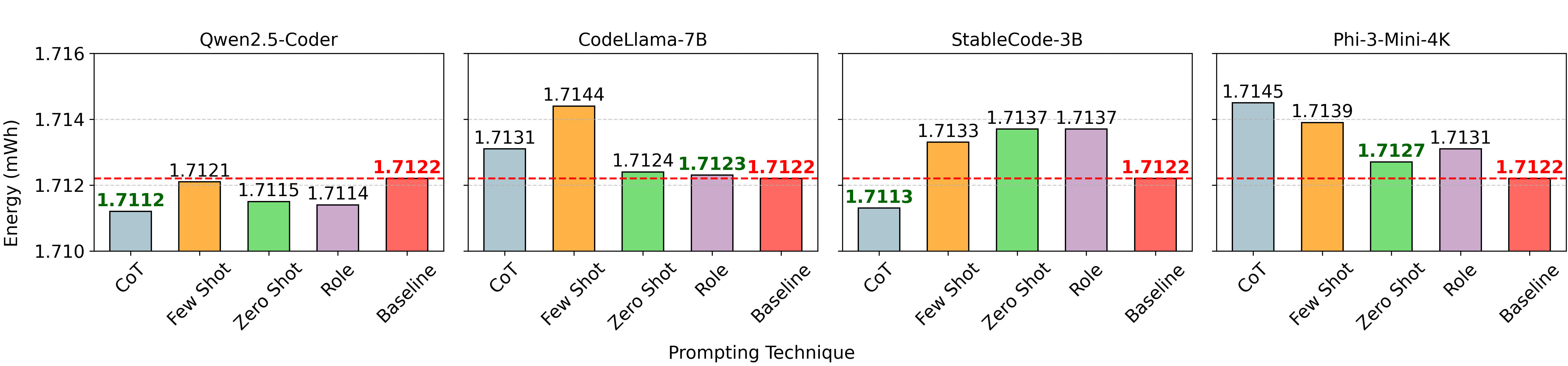}
    \caption{Energy consumption of four SLMs under different prompting strategies compared to the baseline (1.7122 mWh, red dashed line).}
    \label{fig:energycomp}
\end{figure*}

\subsection{Energy Efficiency of SLMs under Different Prompting Strategies}
Fig. \ref{fig:energycomp} shows that improvements in energy efficiency are observed only for Qwen2.5-Coder-3B and StableCode-3B. For Qwen2.5-Coder-3B, all prompting strategies consume less energy than the baseline of 1.7122 mWh, demonstrating consistent benefits from prompting. Among them, CoT prompting performs best at 1.7112 mWh, followed closely by Zero-Shot (1.7115 mWh), Role (1.7114 mWh), and Few-Shot (1.7121 mWh). StableCode-3B, on the other hand, shows improvement only under CoT prompting, where energy consumption drops to 1.7113 mWh, slightly below the baseline, while Few-Shot (1.7133 mWh), Zero-Shot (1.7137 mWh), and Role (1.7137 mWh) all exceed the baseline. For CodeLlama-7B, none of the prompting strategies beat the baseline, but the Role (1.7123 mWh) and Zero-Shot (1.7124 mWh) settings are the most efficient within this model, though still slightly above the baseline, while Few-Shot (1.7144 mWh) is the least efficient. Similarly, Phi-3-Mini-4K does not achieve energy savings against the baseline; however, Zero-Shot prompting is the best option at 1.7127 mWh, followed by Role (1.7131 mWh) and Few-Shot (1.7139 mWh), with CoT consuming the most energy at 1.7145 mWh. Overall, the results demonstrate that Qwen2.5-Coder-3B is the only model consistently more energy-efficient than the baseline across all prompting strategies, while StableCode-3B benefits only from CoT, and the other models fail to outperform the baseline under any prompting method.

\begin{tcolorbox}[colback=gray!10, colframe=black, boxrule=0.3mm, arc=0mm, left=2mm, right=2mm, top=1mm, bottom=1mm]
\textbf{Summary:} Sometimes models fail to benefit from complex prompts such as CoT or Few-Shot, and instead achieve better efficiency with simpler prompts like Zero-Shot or Role. In certain cases, prompting not only fails to improve performance but also increases energy consumption above the baseline.
\end{tcolorbox}

\subsection{Minimum Energy Consumption of SLMs under Optimal Prompting Strategies}
\begin{figure}[t!]
    \centering
    \includegraphics[width=\linewidth]{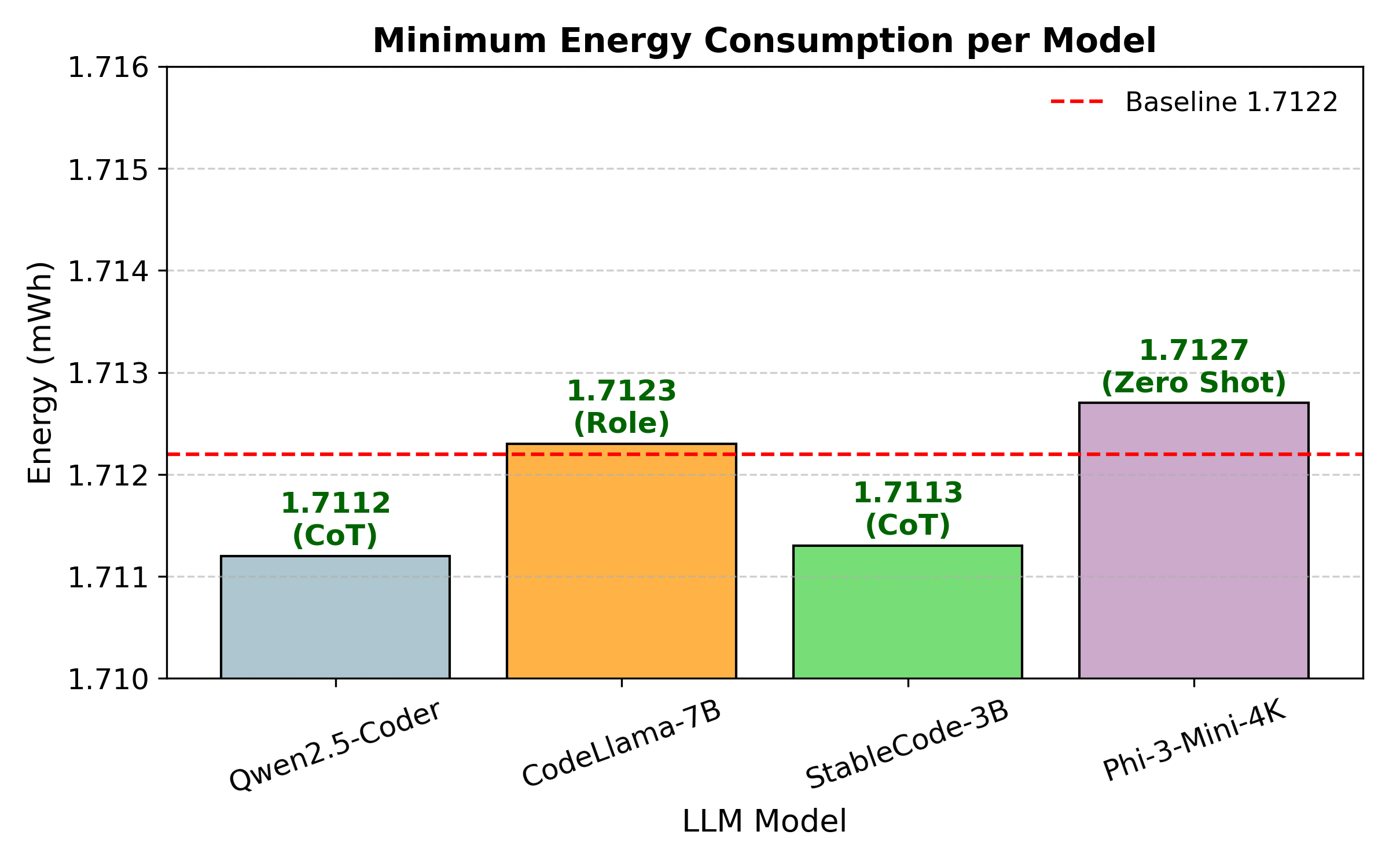}
    \caption{Minimum energy consumption observed for each model under its most efficient prompting strategy, shown in comparison to the baseline (1.7122 mWh)}
    \label{fig:energycomp-2}
\end{figure}

Fig. \ref{fig:energycomp-2} illustrates the minimum energy consumption achieved by each model under its most efficient prompting strategy, compared to the baseline of 1.7122 mWh. Qwen2.5-Coder-3B shows the lowest overall energy usage at 1.7112 mWh with CoT prompting, making it the most energy-efficient configuration. StableCode-3B also benefits from CoT prompting, achieving 1.7113 mWh, slightly below the baseline. In contrast, CodeLlama-7B reaches its minimum with Role prompting at 1.7123 mWh, which is marginally above the baseline, indicating no real energy savings. Similarly, Phi-3-Mini-4K performs best with Zero-Shot prompting at 1.7127 mWh, but this still exceeds the baseline. These results show that prompting can reduce energy consumption for certain models, most notably Qwen2.5-Coder-3B and StableCode-3B, while for others, the best prompting strategy is unable to outperform the baseline.

\begin{tcolorbox}[colback=gray!10, colframe=black, boxrule=0.3mm, arc=0mm, left=2mm, right=2mm, top=1mm, bottom=1mm]
\textbf{Summary:} In summary, CoT works better with models like Qwen2.5-Coder and StableCode-3B, consistently reducing energy consumption below the baseline. Few-Shot prompting did not perform best for any model, as other strategies always achieved lower energy consumption.
\end{tcolorbox}

\subsection{Analysis and Discussion}
The results demonstrate that prompting strategies influence the efficiency of SLMs in different and sometimes unpredictable ways. Among the strategies tested, CoT prompting consistently reduces energy consumption for models like Qwen2.5-Coder and StableCode-3B, outperforming the baseline and confirming that structured reasoning prompts can make certain SLMs more efficient. However, this advantage does not extend to all models. For CodeLlama-7B and Phi-3-Mini-4K, no prompting technique was able to reduce energy consumption below the baseline, with some strategies even leading to higher overhead. This suggests that the relationship between prompting style and energy efficiency is model-dependent, and that improvements in one SLM may not translate to another.

Another important observation is that complex prompting techniques such as Few-Shot do not always provide efficiency gains. In fact, Few-Shot consistently failed to achieve the lowest energy usage for any of the models tested, implying that the additional input complexity can offset potential benefits. Simpler strategies, including Zero-Shot and Role prompting, sometimes deliver better results for certain models, which indicates that prompt simplicity may play an important role in optimizing SLM performance. Overall, these findings highlight that prompt engineering is not a universal solution for achieving efficiency and sustainability in SLMs.

\section{Limitations}
\label{sec:ttv}
This study focuses only on Python, a widely used programming language, so the results may not apply to other languages such as C++ or Java. As a result, the findings cannot be fully generalized across different programming paradigms. In addition, some benchmark problems may have been included in the training data of certain LLMs, which could lead to memorization and an overestimation of their performance. Finally, although the experiments were conducted in a controlled Linux environment, minor system fluctuations and measurement overhead may have introduced small variations in the reported results.

\section{Conclusion}
\label{sec:con}
This study investigated the impact of prompt engineering on the performance and energy efficiency of the code generated by SLMs. The results show that CoT prompting consistently reduced energy use for Qwen2.5-Coder and StableCode-3B, making them the most efficient models. In contrast, CodeLlama-7B and Phi-3-Mini-4K did not perform better than the baseline under any strategy, showing that not all SLMs benefit equally from prompting. Few-Shot prompting also did not give the best result for any model, suggesting that more complex prompts do not always improve efficiency. Overall, the findings show that the success of prompting depends on the model and highlight the need to carefully choose both the model and the prompting strategy to achieve sustainable and energy-efficient code generation.

\bibliographystyle{IEEEtran}
\bibliography{biliography}
\end{document}